\documentclass[runningheads]{llncs}
\usepackage[T1]{fontenc}
\usepackage{graphicx}
\begin{document}

\title{VRSL:Exploring the Comprehensibility of 360-Degree Camera Feeds for Sign Language Communication in Virtual Reality}

\titlerunning{VRSL: Sign Language Communication in Virtual Reality}
\author{Gauri Umesh Rajmane\inst{1} \and
Ziming Li\inst{1}\and
Tae Oh\inst{1} \and
Roshan Peiris\inst{1}}
\authorrunning{Rajmane et al.}
\institute{Rochester Institue of Technology, Rochester NY 14623, USA\\ \email{gr3504, zl1398, thoics, rxpics@rit.edu}\\
}
\maketitle

\begin{abstract}
This study explores integrating sign language into virtual reality (VR) by examining the comprehensibility and user experience of viewing American Sign Language (ASL) videos captured with body-mounted 360-degree cameras. Ten participants identified ASL signs from videos recorded at three body-mounted positions: head, shoulder, and chest. Results showed the shoulder-mounted camera achieved the highest accuracy (85\%), though differences between positions were not statistically significant. Participants noted that peripheral distortion in 360-degree videos impacted clarity, highlighting areas for improvement. Despite challenges, the overall comprehension success rate of 83.3\% demonstrates the potential of video-based ASL communication in VR. Feedback emphasized the need to refine camera angles, reduce distortion, and explore alternative mounting positions. Participants expressed a preference for signing over text-based communication in VR, highlighting the importance of developing this approach to enhance accessibility and collaboration for Deaf and Hard of Hearing (DHH) users in virtual environments.
\end{abstract}

\section{Introduction}
Virtual reality (VR) applications, such as games, can immerse users completely in virtual environments by engaging the senses of sight, hearing, and touch ~\cite{djain}. Among these, sound plays a critical role in conveying spatial, interaction, and contextual cues, enhancing the realism and responsiveness of VR experiences ~\cite{Li},~\cite{soundHapticVR},~\cite{djain}. In collaborative virtual spaces, such as multiplayer games and social VR, sound becomes even more pivotal, as effective communication and teamwork are integral to user experience ~\cite{Luna}. Verbal interactions, enabled through voice chat, and non-verbal audio cues, such as spatially localized sounds or action-specific noises, help participants coordinate and foster a sense of presence within the shared virtual environment.

The immersive VR experience often impacts Deaf and Hard of Hearing (DHH) users due to the inaccessibility of auditory information and the limitations of existing communication methods in VR. Despite existing accessibility solutions such as closed captioning, chat rooms, and sign language communication through hand-tracking and avatars, these methods present notable limitations. Closed captioning, while useful, primarily facilitates communication between hearing and nonhearing individuals and is ineffective for DHH-to-DHH interactions~\cite{Kim}. Hand-tracking and avatar-based sign language face significant technical hurdles, including continuous hand posture detection and resource-heavy rendering processes, resulting in inaccurate and poorly rendered signs ~\cite{Anderton}. Text-based chat rooms lack non-verbal communication, which is vital for DHH users, as written English and captions fail to offer equivalent access, functionality, or expressive power for signers.\cite{Ang},~\cite{Kushalnagar},~\cite{Gokcay}.

Our study addresses communication challenges in virtual reality (VR) by exploring the integration of sign language video feeds using 360-degree body-mounted cameras to capture and convey American Sign Language (ASL). The use of 360-degree video provides comprehensive capture of signing movements and facial expressions where necessary to accommodate the spatial and directional nuances of ASL and enhancing its clarity in VR environments. This approach aims to enable real-time ASL interaction, offering an accessible alternative to text- or voice-based communication.

To evaluate this method, we assess the clarity and user experience of ASL signs recorded from three body-mounted camera positions: head, shoulder, and chest. While the primary focus is on the comprehensibility of these recordings, the findings lay the groundwork for facilitating future research on real-time interaction and collaboration in VR. Insights from this research inform the refinement of camera setups and pave the way for more inclusive virtual environments through innovative video-based communication solutions.

Thus, our research addresses the following primary research questions, 
\begin{itemize}
    \item {RQ1}: Can video feeds from body-mounted 360-degree cameras accurately capture and convey ASL in VR?
    
    \item {RQ2}: Which body mount positions optimize the clarity and visibility of ASL signs?

\item{RQ3}: What are the potential advantages and limitations of this method compared to existing VR communication tools?

\end{itemize}

By addressing these questions, this research aims to establish the viability of 360-degree video feeds as a precursor to more robust and natural methods of DHH communication in VR.

\section{Prior Work}
\subsection{Communication in Virtual Reality}
Communication in virtual reality[VR] has been a research interest since the ’90s ~\cite{Biocca}. This interest continues to grow among researchers in the fields of technology, human-computer interaction, and psychology. The reason behind this interest is the uniqueness of communication in virtual reality. What differentiates VR from other computer-mediated communication(CMC) mediums is that it has the potential to integrate the full spectrum of verbal and non-verbal communication~\cite{Dzardanova}. Dzardanova et al. indicated that VR can be as rich and intricate as interpersonal face-to-face interactions ~\cite{Dzardanova}.

Communication in VR is important, especially in the context of social VR and multiplayer gaming.  In both of these contexts, collaboration and engagement with other users is a core part of the experience~\cite{Wei}. Prior research shows us that the quality and frequency of communication among players influence the quality of collaboration in immersive gaming which ultimately affects the user experience~\cite{Luna}

\subsection{DHH Experience in VR Communication}
A 2022 systematic literature review on communication in immersive social VR, filtered through 1058 papers and took in account 32 for a detailed analysis - to understand the factors that affect communication and how to evaluate communication in VR~\cite{Wei}. Of all these, one research study focused on DHH users. We see that there is a lack of research that aims at understanding deaf users' experience in virtual reality. While notable research exists on sound accessibility methods, and hearing-DHH communication, highlighted in sections 2.0.3, 2.0.4, and 2.0.5, studies on DHH-DHH communication remain limited.

A notable research gap exists in our understanding of communication specifically between two DHH users within the context of VR~\cite{Luna}. Furthermore, insights into the experiences of DHH users in virtual reality itself remain limited.

\subsection{Sign Language and Collaboration in VR}
Previous research underscores the significance of communication in fostering collaboration within multiplayer immersive games~\cite{Luna}. Luna et al. discovered that over half of their DHH participants identified collaboration as the highlight of gameplay, emphasizing the pivotal role of communication in determining collaboration levels~\cite{Luna}. Their study revealed a preference among DHH participants for employing various communication modes while collaborating. Proficient ASL users, uncertain of others' proficiency, and hearing participants not adept in ASL resorted to inventing signs or using gestures, even when text-based communication was available~\cite{Luna}. This suggests that text communication may not always be the optimal or the preferred mode of communication. The findings from this study inspire my study to incorporate non-verbal and ASL communication in VR.

Prior research also highlights critical barriers to effective collaboration in Virtual Reality (VR) environments, particularly for multi-party conversations  ~\cite{Anna}. The paper ~\cite{Anna} highlights the need to create social presence, establishing clear conversational roles (such as speaker, bystander etc.) and the need for shared visual references.

An additional illustration comes from users in VR Chat, where made-up signs and gestures were employed due to improper rendering~\cite{GameRevolution}. The challenges of capturing signing in VR were elucidated in a 2022 study by Quandt et al., emphasizing the resource-heavy nature of real-time avatar creation and hardware limitations in headsets, resulting in significant blind spots and a restricted field of view for sign sensing~\cite{Quandt}. This limitation hinders the recognition of crucial visual and spatial parameters in ASL, including hand shapes, physical location, and facial expressions, as identified by Friedman~\cite{Friedman}. Research in linguistics reinforces the importance of the five parameters—hand shape, palm orientation, movement, location, and expression/non-manual signals—for effective ASL communication{~\cite{Friedman}}. These parameters are essential and set a foundation for testing the clarity of ASL communication in my study.

In the broader realm of communication and technology, studies explore ASL communication through virtual interpreters {~\cite{Berke}}. Berke et al. test a hands-free prototype of a body-mounted camera and speaker to allow for ASL communication between a DHH user and a hearing user mediated by an interpreter{~\cite{Berke}}. The interpreter would view the DHH user signing via the body-mounted camera and translate the information into audio which would be heard via the speaker. Chat in the Hat demonstrates that a solution via body-mounted cameras to capture a user signing could be feasible and motivates efforts to use body-mounted cameras in our study{~\cite{Berke}}.

\subsubsection{Closed Captioning}
Closed captioning ~\cite{Kim} and VR as an assistive tool are other solutions in the intersection of communication, DHH users, and technology, but the primary objective is to bridge the communication gap between hearing and deaf users~\cite{Miller},~\cite{Batas}. Closed captioning works by translating audio to text and vice versa{~\cite{Kim}}. Although this method is useful, it is aimed to bridge the communication gap between a hearing and a non-hearing person. This solution cannot be used among two DHH users. We see a lack of research and solutions that cater to DHH-to-DHH communication.

\subsubsection{Sound Accessibility}
Prior research and accessibility methods look at using haptic and visual cues to make sound accessible to users( ~\cite{JainChiu}, ~\cite{SoundModVR}, ~\cite{Ziming} Sound viz VR visualizes sound characteristics and sound types for several types of sounds in VR experience{~\cite{Li}}. Similarly, Towards sound accessibility built a prototype to provide a  visual and haptic substitute for VR sounds{~\cite{djain}}. While such methods do help to provide context for sounds in virtual environments, they cannot be used for communication among players in VR or for fast-paced games. 

\subsubsection{Hand-Tracking and Sign Language Recognition}
Sign Language in VR is currently implemented by using a combination of hand-tracking and avatars. The controllers and headset need to accurately capture the signer's hand movements and then translate that as a signing avatar. We see hardware limitations in accurately capturing signing movement, as headsets have significant blind spots{~\cite{Quandt}. Signs that involve touching face or body parts cannot be captured accurately by the cameras in VR headsets{~\cite{Quandt}}.

Prior research explored the role of avatars in supporting non-verbal communication in VR ~\cite{Smith}, ~\cite{Anna}.  Highlighting that detailed and realistic avatars enhance co-presence and emotional engagement~\cite{Anna}, ~\cite{Smith}. 

Technical limitations on rendering avatars exist even with state-of-the-art controllers and programs. Sign language requires some very precise movements, but not all controllers are capable of capturing and rendering it quickly in real-time{~\cite{Quandt}}. Creating avatars is very resource heavy and animating signs becomes a burden on the VR platform{~\cite{Quandt}}. We see an example in a popular social VR application VRChat, where DHH users had to invent their own signs to communicate in VRChat, due to the avatars' signs being rendered inaccurately and poorly{~\cite{GameRevolution}}. Additionally, not all VR applications have avatars.

\subsubsection{Text-Based Communication}
Chat rooms are another useful communication option. This text-based communication medium cannot incorporate any non-verbal communication. They fail to provide DHH users with communication capabilities comparable to the spoken conversation hearing users readily and quickly employ. Prior research shows that deaf signers commonly utilize sign language interpreters as written English or captioning does not provide the same access, functional equivalency, or expressive power for signers \cite{Ang},~\cite{Kushalnagar}. We may also consider that prelingually deaf people may feel more comfortable communicating with sign language as sign language is usually their first language and English their second{~\cite{Cormier}}.

\section{Methodology}

To address our research questions, we designed a study with the aim of evaluating the comprehensibility of signs presented in 360-degree videos in VR. The VR application that presented the 360 video and evaluated their comprehensibility, was developed using Unity 3D and deployed on a Meta Quest headset, allowing participants to view ASL videos and select the correct interpretation from multiple-choice options. 

\subsection{Study Design}
This study used a design within the subject with one independent variable: location of the camera mounted on the signer, which had three levels as conditions (head, shoulder, and chest). The dependent variable was the accuracy of ASL identification. Perceived effort was measured using a NASA TLX questionnaire~\cite{tlx} and any additional subjective feedback was collected using an exit survey questionnaire.

The study utilized three main task categories: minimal pairs, non-minimal pairs, and sentences. These categories are based on the widely accepted framework of ASL linguistic parameters — hand-shape, palm orientation, movement, location, and non-manual signals{\cite{Gu},~\cite{Friedman}}. 

\textit{Minimal Pairs}: Minimal pairs consist of signs that differ by only one parameter (e.g "Please" and "Sorry" are a hand-shape minimal pair, differing only in hand shape while palm orientation, movement, location and non-manual signals remain same){\cite{Gu}}. These tasks were included to evaluate whether any ASL parameter is more prone to errors.

\textit{Non-Minimal Pairs:} Since most everyday interactions may not involve minimal pairs, non-minimal pairs were included to evaluate the system's effectiveness in general comprehensibility to single ASL words. 

\textit{Sentences and Contextual Interpretation:} Sentences were included to examine how context influences sign interpretation, as meaning is often clearer within sentences than in isolated words {\cite{Berke}}. This assessment helps determine whether context enhances accuracy in sign recognition and comprehension. 

Each condition comprised 10 tasks, totaling 30 tasks per participant. The tasks and conditions were randomized to prevent order bias and were balanced for difficulty across minimal pairs (5 tasks), non-minimal pairs (2 tasks), and sentences (3 tasks), ensuring equal representation across different camera angles.

\subsection{Procedure}
Participants signed a consent form and filled out a demographic questionnaire about their VR experience and its collaborative use. They received information about the study and procedure via e-mail. After putting on the VR headset, they received instructions on using the controllers and a walkthrough of the VR application through textual and visual information in the VR environment. During the VR task, participants identified the presented ASL word/sentence. Between each condition, they removed the headset, completed the TLX survey, and provided subjective feedback on a form. At the end, they filled out an exit survey for qualitative feedback and overall preferences between the three camera locations. The complete study took approximately 20-30 mins to complete.

\subsection{Participants}
We recruited 10 DHH participants from the institution (6 females, 4 males, ages 18-55). The group included 9 deaf or hard-of-hearing participants and one hearing participant fluent in ASL. All participants had previous experience with VR devices; 4 had used VR collaboratively, while 6 had not. Participants were recruited through flyers and word-of-mouth at the institution and were paid \$20 for completing the experiment.

\subsection{Apparatus: Video Capture and Camera Mounting Positions}

To explore the feasibility of this method, our study utilized recorded 360-degree videos rather than live-streamed content to simulate video-based sign language communication. To capture the ASL 360-degree videos, we used a Ricoh Theta V camera, which records 360-degree dual fish-eye video. The dual fish-eye video streams were processed and converted to an equirectangular format in Adobe After Effects for the user study.

The camera was mounted on a DHH signer using body-mounting accessory straps, allowing placements on the head, shoulder, chest, and belt. The signer received a list of words and phrases to sign before recording at a lab facility. The signer wore a VR headset during recordings to simulate another VR player signing to them.  

To explore the comprehensibility of ASL using body-mounted 360-degree cameras in VR, careful consideration was given to camera placement. The chosen positions—head, shoulder, and chest—were informed by existing research, technical feasibility, and practical constraints. The following mounting positions were explored:

%Initial attempts to mount the camera under the VR headset using adhesive tape failed due to the weight of the camera and insufficient adhesive strength. The camera would dislodge during head movements, leading to unreliable recordings. Consequently, the head-mounted camera was secured using body-mounting straps attached to the headset, ensuring stability during recording.

%Lighter camera technology or improved mounting mechanisms, such as adhesive strips designed for VR headsets, under-headset mounting could be explored. 

\begin{figure}
    \centering
    \includegraphics[width=\linewidth]{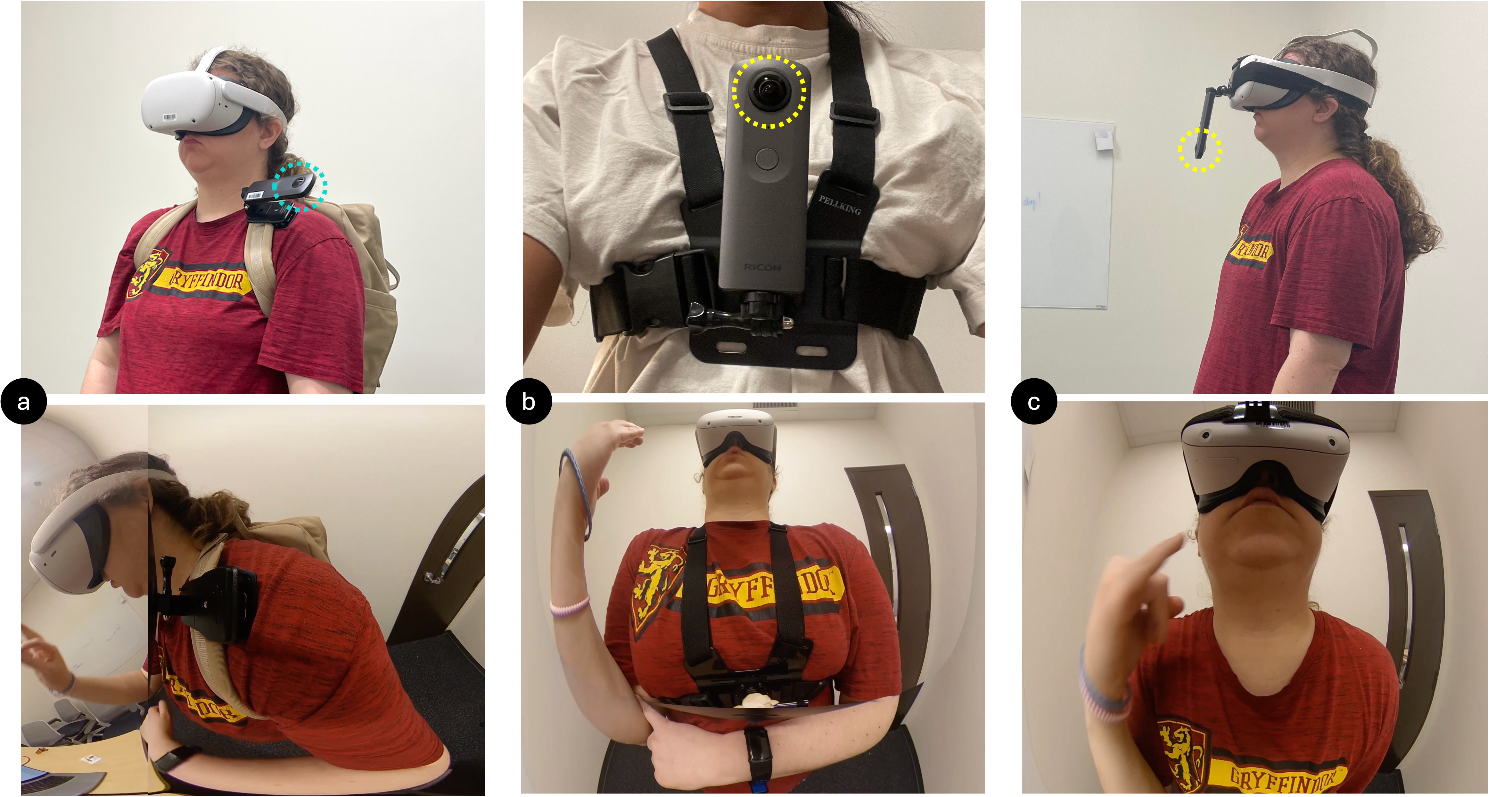}
    \caption{Camera mounting positions and their view. Encircled are the location of the camera. Top row shows the shoulder-mounted (a-top), chest-mounted (b-top) and head-mounted cameras (c-top) on the signer. Bottom row shows the footage captured by the shoulder (a-bottom), chest (b-bottom) and head (c-bottom) mounted cameras }
    \label{fig:cameraPositions}
    %\Description{This image contains two rows of photos depicting different camera mount positions for virtual reality (VR) tasks and their corresponding perspectives. The top row shows three individuals wearing VR headsets, each with a distinct camera mount configuration: (a) a camera attached to the shoulder via a backpack strap, (b) a camera mounted to the chest using a harness, and (c) a camera mounted on the head via the VR headset itself. The bottom row provides views of how the VR environment is captured from each mount's perspective, highlighting variations in perspective distortion for the shoulder, chest, and head positions.}
\end{figure}

% \begin{figure}
%     \centering
%     \includegraphics[width=0.5\linewidth]{images/Head Mounting on Signer.jpeg}
%     \caption{Head-Mounted Camera on Signer}
%     \label{fig:enter-label}
% \end{figure}

% \begin{figure}
%     \centering
%     \includegraphics[width=0.5\linewidth]{images/Video Footage from Head mounted.png}
%     \caption{Head-Mounted Camera Footage of Signer performing ASL}
%     \label{fig:enter-label}
% \end{figure}
\subsubsection{Shoulder-Mounted Position}
The shoulder-mounted position (Fig.~\ref{fig:cameraPositions}(a)) provides a stable, slightly angled perspective that closely approximates a natural third-person viewpoint for observing signing. Stability is crucial for maintaining clarity during dynamic movements in VR. Shoulder mounting ensured that the dominant hand’s movements were visible. However, it posed challenges for signs performed on the opposite side of the body, limiting the camera’s ability to capture complete gestures.

% \begin{figure}
%     \centering
%     \includegraphics[width=0.5\linewidth]{images/Shoulder Mounting on Signer.png}
%     \caption{Shoulder Mounted Camera on Signer}
%     \label{fig:enter-label}
% \end{figure}

% \begin{figure}
%     \centering
%     \includegraphics[width=0.75\linewidth]{images/Video footage from Shoulder Mounted.png}
%     \caption{Shoulder Mounted Camera Footage of Signer performing ASL}
%     \label{fig:enter-label}
% \end{figure}

\subsubsection{ Chest-Mounted Position}
The chest-mounted position (Fig.~\ref{fig:cameraPositions}(b)) was included due to its central placement, offering an unobstructed view of both hands during signing. This position was hypothesized to provide a balanced perspective, making it well-suited for interpreting complex ASL gestures. Furthermore, a wearable, necklace-style camera could feasibly replicate this angle, presenting a practical option for integration in future designs.

\subsubsection{Head-Mounted Position}

The head-mounted position (Fig.~\ref{fig:cameraPositions}(c)) was selected as it offers a vantage point similar to the signer’s perspective. Prior research, such as the "Chat in the Hat" study~\cite{Berke}, identified this position as providing high accuracy for sign interpretation due to its alignment with natural head movements. Additionally, many VR headsets are already equipped with cameras positioned on the headset, offering a similar viewing angle. This suggests the potential for integrating cameras in future headsets at comparable positions, should this approach prove feasible.

% \begin{figure}
%     \centering
%     \includegraphics[width=0.5\linewidth]{images/Chest Mounting on Signer.jpg}
%     \caption{Chest Mounted Camera on Signer}
%     \label{fig:enter-label}
% \end{figure}

% \begin{figure}
%     \centering
%     \includegraphics[width=0.5\linewidth]{images/Video footage from Chest Mounted.png}
%     \caption{Chest Mounted Camera Footage of Signer performing ASL}
%     \label{fig:enter-label}
% \end{figure}

%\subsubsection{Other Considered Positions (Excluded):}

% \textbf{Wrist-Mounted Position:}
% Rejected due to excessive movement during signing, which may result in a blurred or obstructed view of the signer’s hands.
% \\
% \textbf{Belt-Mounted Position:}
% Rejected due to limited visibility of upper-body gestures and facial expressions, which are critical for interpreting ASL accurately.

\subsection{Software and Setup Design}
The videos were presented within a VR application developed using Unity and deployed on a Meta Quest headset. For each condition, participants viewed a video and selected one of three options: the correct ASL interpretation, a similar but incorrect option, or an option to skip the question. The 'skip' option was included to minimize random guessing. Participants were allowed to replay the videos as many times as needed, with no time constraints. Each participant interpreted a total of 30 videos, with 10 videos corresponding to each camera position. The order of questions and camera conditions was randomized to reduce potential bias.

The study software included the following steps to guide the participant through onboarding and the setup. 

\begin{enumerate}
    \item {On-boarding Experience (Fig.~\ref{fig:screenshots}(a)-(b)):} Participants were guided through an on-boarding process within the VR environment. Instructions on navigating the application, using the controller, and selecting options were provided through a combination of images and videos .
    \item {Video Viewing Customization (Fig.~\ref{fig:screenshots}(c)-(d):} Participants had the option to customize the location of the video on their screen, choosing from bottom-left, top-left, bottom-right, top-right, or a fixed position. This feature allowed for flexibility in viewing and offered valuable insights into individual user preferences.
    \item {Video Box Size:} Multiple video box sizes were tested to ensure they were large enough for clear viewing but small enough to avoid obstructing the VR environment. Users could also choose whether the video box was fixed in one location or moved dynamically with their headset.
    \item     {Task Question(Fig.~\ref{fig:screenshots}(e):} For each task, the participant watched a video of the signer and chose the correct option from two given choices.(Fig.~\ref{fig:screenshots}(e) the choices are "Calculator" and "Table)

\end{enumerate}

% \begin{figure}
%     \centering
%     \includegraphics[width=0.75\linewidth]{images/Overview.png}
%     \caption{ VR Software design with participant on-boarding and instructions in VR environment before starting the task.}
%     \label{fig:Figure 1}
% \end{figure}

% \begin{figure}
%     \centering
%     \includegraphics[width=0.75\linewidth]{images/Instructions.png}
%     \caption{ VR Software design with participant on-boarding and instructions on how to use the controller settings and navigate.}
%     \label{fig:Figure 1}
% \end{figure}

% \begin{figure}
%     \centering
%     \includegraphics[width=0.75\linewidth]{images/Video Viewing Options.png}
%     \caption{ VR Software design with video location options for participants. }
%     \label{fig:Figure 2}
% \end{figure}

% \begin{figure}
%     \centering
%     \includegraphics[width=0.75\linewidth]{images/Video Box Preview.png}
%     \caption{VR Software design with video location option demo for participant to try before starting task. }
%     \label{fig:Figure 3}
% \end{figure}
% \begin{figure}
%     \centering
%     \includegraphics[width=0.75\linewidth]{images/Shoulder-Mounted-Question.png}
%     \caption{Task in VR software showing the video in the bottom-right viewing position. The video displays the ASL for "Calculator" from the shoulder-mounted camera condition, with options to select the correct interpretation, an incorrect option, or to skip the question.}
%     \label{fig:Figure 4}
% \end{figure}

\begin{figure}
    \centering
    \includegraphics[width=\linewidth]{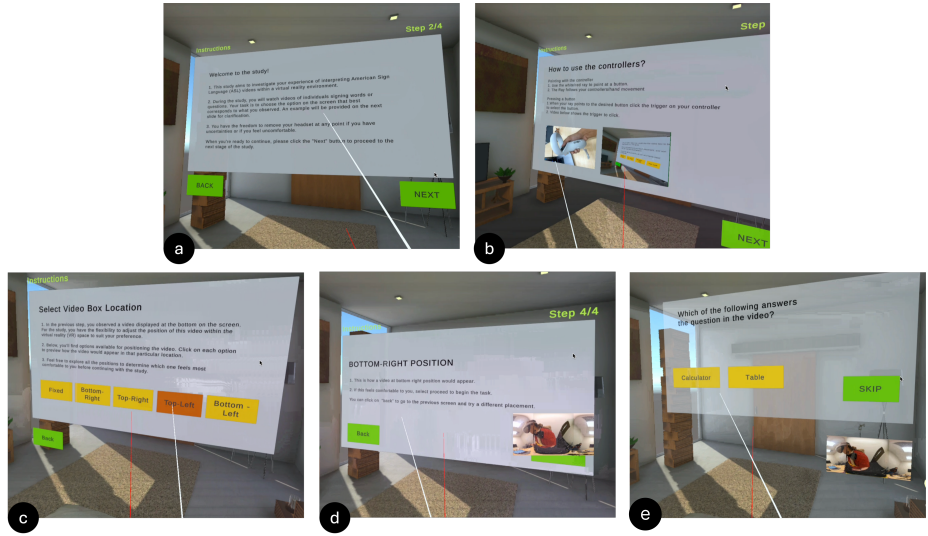}
    \caption{Screenshots of the VR study software: (a) Welcome screen in the VR Application (b) participant on-boarding and instructions on how to use the controller settings and navigate. (c) Options for user to select where they want To view the video, (d) Demo for the user to see a bottom-right video viewing position, (e) Task for user to interpret video being signed for the word "calculator"}
    \label{fig:screenshots}
    %\Description{This image contains a sequence of screenshots illustrating the VR interface and study workflow. The first screenshot (a) introduces the study goals, displaying a welcome screen with instructions. The second (b) provides guidance on using the VR controllers for navigation and selection. The third (c) allows participants to choose the video box location in the VR environment, with options like bottom-right, top-left, or fixed positions. The fourth (d) shows a preview of the selected video box placement, helping participants visualize its appearance within the virtual space. The final screenshot (e) presents a question interface prompting participants to answer based on the video content, with selectable options like "Calculator" or "Table" displayed as buttons.}
\end{figure}

\section{Results}

Figure~\ref{fig:results}(a) shows the overall comprehensibility based on the recognition accuracy of the sign-language phrases presented from the different camera positions. As observed, while the recognition accuracies for Head and Shoulder mounted angle resulted in higher accuracies (85\% and 84\% respectively) than Chest mounted angle, a Repeated Measures ANOVA yields no significant differences ($P>0.05$) . Overall, participants correctly identified an average of 25 out of 30 ASL tasks, achieving a 83.3\% comprehension success rate across all conditions. Among the three conditions, the shoulder-mounted camera had the highest accuracy. Participants correctly answered an average of 85\% of the questions for the should-mounted condition. 

Analysis of minimal pair accuracy across different linguistic components provides further insight into participant performance. Participants achieved the highest accuracy for location minimal pairs, with a perfect score of 100\% (30 out of 30 correct). Movement minimal pairs followed closely with 96.67\% accuracy, while handshape and expression minimal pairs both achieved 90\% accuracy. Palm/orientation minimal pairs had the lowest accuracy at 80\%, indicating a relative difficulty in distinguishing these features. For single words, participants demonstrated a high accuracy rate of 96.67\%. However, in more complex tasks involving understanding a sentence, accuracy dropped to 70\%, suggesting increased cognitive demands in these scenarios.

Figure~\ref{fig:results}(b) shows that TLX data indicated slightly more favorable scores for the shoulder mounted angle compared to the chest and head mounted angles. When each participant was inquired about the preferred angle, no strong preference emerged for any condition, with 4 participants preferring the chest mount, 4 preferring the shoulder mount, and 2 having no specific preference. However, six out of ten participants indicated they preferred signing via our proposed method over texting or chatting in VR. No strong preference for video viewing location emerged, with selections varying according to personal preference.  

%Each study session took between 12-18 minutes per participant, including instructions, tasks, and surveys. Each condition (10 tasks per condition) took between 2-5 minutes to complete, with less than a minute per task.

\begin{figure}
    \centering
    \includegraphics[width=\linewidth]{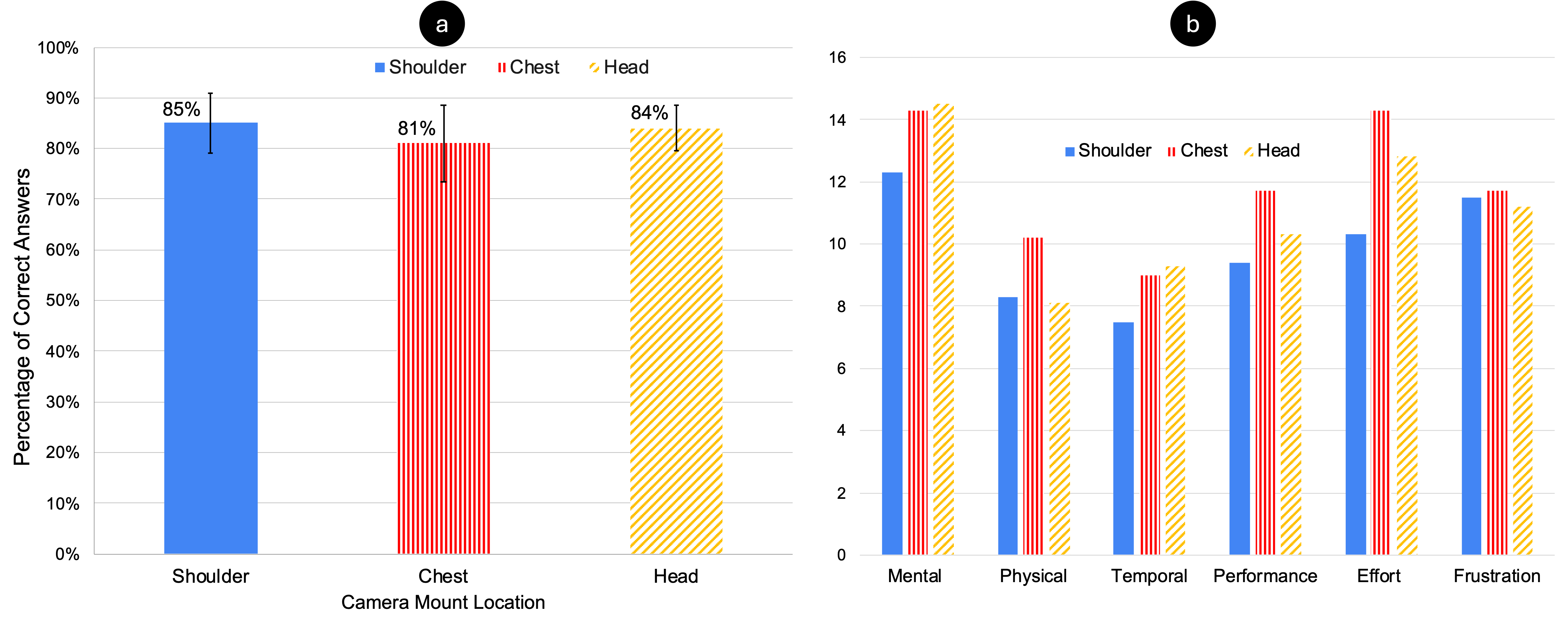}
    \caption{(a): Average accuracy percentage and  (b): TLX scores across mental, physical, temporal, Success, Effort and Frustration for the three conditions}
    \label{fig:results}
    %\Description{This image contains two graphs summarising performance and TLX workload data for different camera mount locations. The first graph (Graph A) is a bar chart comparing the percentage of correct answers across three camera mount configurations: shoulder, chest, and head. It shows slight variations in accuracy, with the shoulder mount achieving 85\%, the head mount 84\%, and the chest mount 81\%. The second graph (Graph B) displays subjective workload ratings (mental, physical, temporal, performance, effort, and frustration) for the same configurations. The head and chest mounts generally show higher workload ratings across most categories, while the shoulder mount has relatively lower scores, suggesting less effort and frustration.}
\end{figure}

% \begin{figure}
%     \centering
%     \includegraphics[width=0.75\linewidth]{images/Screenshot 2024-06-24 at 9.49.53 AM.png}
%     \caption{Average comprehensibility accuracy for conditions head, shoulder and chest}
%     \label{fig:Figure 5}
% \end{figure}

% \begin{table}[ht]
% 	\centering
%        %% \begin{adjustbox}{width=1\textwidth}
%         \small
% 	\caption{One-way Repeated Measures ANOVA analysis for comprehensibility accuracy}
% 	\begin{tabular}{lrrrrr}
% 			\toprule
% 			Cases & Sum of Squares & df & Mean Square & F & p  \\
% 			\cmidrule[0.4pt]{1-6}
% 			Camera Location & $0.867$ & $2$ & $0.433$ & $0.255$ & $0.776$  \\
% 			Residuals & $45.800$ & $27$ & $1.696$ & \\
% 			\bottomrule
% 			% \addlinespace[1ex]
% 			% \multicolumn{6}{p{0.5\linewidth}}{\textit{Note.} Type III Sum of Squares} \\
% 	\end{tabular}
%       %%  \end{adjustbox}

%       \label{tab:anova}
% \end{table}

% \begin{figure}
%     \centering
%     \includegraphics[width=1\linewidth]{images/TLX .png}
%     \caption{NASA TLX scores for mental demand, physical demand, temporal demand , success, effort and frustration across head (yellow), chest (blue) and shoulder (red) conditions. Lower scores indicate a lower task load.}
%     \label{fig:Figure 6}
% \end{figure}

\section{Discussions}
In addressing RQ1 and RQ2, the results indicate that the shoulder-mounted camera provided the highest accuracy for ASL identification, with an 85\% comprehension success rate. However, the difference was not statistically significant. This may suggest that the shoulder position offers a slightly better perspective for interpreting ASL compared to head and chest mounts. The TLX scores also slightly favored the shoulder mount, indicating that participants found this position somewhat less effortful. However, the lack of a significant difference between conditions implies that all three positions may be viable for ASL communication in VR.

Addressing RQ3, from the subjective feedback of participants, we found that distortion introduced by the 360-degree cameras was a significant factor affecting the clarity of the signs. Particularly signs performed at the periphery of the field of view were affected most by the distortion. This made it challenging to accurately interpret finger movements and shapes essential for distinguishing minimal pairs in ASL. This might explain why the chest mounted location was the least preferred condition, as it is subject to the most distortions, whereas the shoulder mount experienced the least.

The shoulder condition had a drawback, in that it favored one side of the body where the camera was positioned, making it difficult to view signs performed on the opposite side (the side was selected based on the handedness of the signer). P3 reported, \textit{“It was hard to see the sign when it was on the other side, like sunset or sunrise or signs from one side to the other. I only caught half of it.”} They also recommended considering a non-body-mounted camera or a camera mounted in the space where the VR user was, which would provide a third-person perspective that feels more natural compared to a body-mounted first-person perspective. P5 mentioned, \textit{“I think a more distant camera would help with the distortion of the person; right now it's just way too close up.”}

Another theme from participant feedback was the unnatural viewing angles presented by a body-mounted camera, especially the head-mounted camera. This impacted the natural look of the signing, making it harder for participants to follow the signs accurately. P6 reported, \textit{“It was more difficult with a non-favorable or unnatural position to watch from an upper or bird's-eye view, plus it was hard seeing fingerspelling.”} All body-mounted positions differ from how individuals usually see signs, which is facing the other participant. Further work is needed to make the camera angles and video viewing more natural, as P4 indicated, \textit{“The video needs to look and feel more natural - the video hurt my eyes a bit, especially when I was trying to decipher the signs.}”

Analysis of minimal pair performance showed that while participants achieved high accuracy overall, certain features such as palm/orientation pairs posed more difficulty compared to other categories like location and movement. The lower 80\% accuracy in palm/orientation minimal pairs suggests that finer details in sign articulation are more challenging to discern due to camera positioning, distortion effects, and the nuanced hand rotations that require precise visual interpretation. Distortion from the 360-degree cameras, particularly in the chest-mounted position, may have further contributed to the difficulty in perceiving these subtle hand orientations. In contrast, the 100\% accuracy in location minimal pairs indicates that spatial elements of ASL are better captured across conditions, as larger, more pronounced movements are less affected by distortion and easier to identify. The relatively high accuracy in movement minimal pairs (96.67\%) suggests that dynamic gestures were generally well perceived, likely due to their larger scale and more distinguishable motion patterns. Additionally, the lower performance in sentence-based tasks, with a 70\% accuracy, suggests that contextual understanding requires further improvements in video clarity, positioning, and possibly better framing to ensure that the entire sign sequence is visible without obstruction. This performance gap also indicates that cognitive load increases when participants must rely on sequential recognition rather than isolated sign features.

The average comprehension success rate of 83.3\% across all conditions shows the potential effectiveness of  video feeds for ASL communication in VR. Participants' mixed preferences for camera mounting positions and video viewing locations suggest that customization options could enhance the user experience by accommodating individual preferences. The study duration and task completion times were efficient, with most tasks taking less than a minute to complete. This indicates that our method has the potential to be effective and practical for real-world applications with further refinement. The preference for signing over texting or chatting in VR among most participants highlights the value of this communication method for the DHH community.

\section{Limitations and Future Works}

Our study's findings highlight several areas for future research and development. One major drawback identified was the distortion caused by the 360-degree cameras, particularly for signs performed at the edges of the field of view. This issue was especially pronounced in the chest-mounted condition, which participants found least clear due to the increased distortion. The shoulder-mounted camera, while more accurate, had limitations in providing a clear view of signs performed on the opposite side of the body.

Participants also noted that the unnatural viewing angles from body-mounted cameras, especially the head-mounted camera, made it difficult to follow the signs accurately. Future work would focus on experimenting with different cameras to reduce distortion and consider non-body-mounted cameras to provide a more natural third-person perspective. Exploring non-fish-eye or multi-camera setups could also help in reducing distortion and providing a clearer viewing experience.

Our primary aim was to explore the comprehensibility of video feeds in VR, but a limitation of our approach is the lack of real-time communication. Testing  real-time collaboration would mimic a more realistic scenario and uncover collaboration challenges and requirements. Experiments with real-time feedback mechanisms within the VR environment could provide us with more insights into the applications of this prototype.

Moreover, future studies should include a larger and more diverse participant group to provide more comprehensive insights into user preferences and system performance. Investigating alternative mounting positions or configurations that better replicate natural signing interactions could further enhance the effectiveness of this technology.

Additionally, complementary to the proposed approaches, we aim to explore integrating other modalities such as multimodal captions~\cite{cap1},~\cite{cap2},~\cite{cap3},~\cite{cap4}, etc. to provide alternative communication options in VR for DHH users.

\section{Conclusion}
In conclusion, while body-mounted 360-degree cameras show potential for capturing and conveying ASL in VR, improvements are needed to address issues related to distortion, field of view, and the natural look of signing. By exploring alternative technologies and configurations, future research can build on these findings to develop more effective solutions for ASL communication in virtual environments.

\bibliographystyle{splncs04}
\bibliography{sample-base.bib}

%\appendix

\end{document}